\documentclass{JFM-FLM_Au}
\usepackage{enumitem}
\usepackage{xcolor}
\usepackage{ulem}
\usepackage{siunitx} 
\usepackage{todonotes}
\usepackage{amsmath} 
\usepackage{makecell}
\usepackage{tabularray}
\usepackage{lscape}
\usepackage{booktabs}
\usepackage{multirow}
\usepackage{array}
\usepackage{longtable}

\lefttitle{Jing Dong et al.}
\righttitle{Journal of Fluid Mechanics}

\title{An experimental study on the heat transport in porous media convection}

\author{Jing Dong\aff{1}, Lu Zhang$^*$\aff{1,2} \and Ke-Qing Xia$^*$\aff{1,2}}

\affiliation{\aff{1}Department of Mechanics and Aerospace Engineering, Southern University of Science and Technology, Shenzhen, China
\aff{2} Centre of Complex Flow and Soft Matter, Southern University of Science and Technology, Shenzhen, China}

\corresau{Lu Zhang, \email{zhangl39@sustech.edu.cn}; Ke-Qing Xia \email{xiakq@sustech.edu.cn}}

\begin{document}
\maketitle
\begin{abstract}
We investigate the heat transport in porous media convection over a wide Rayleigh--Darcy number range of $26.8\leq Ra\leq 2.62\times 10^5$, and a Darcy number range of $6.18\times10^{-7}\leq Da\leq 1.21\times 10^{-5}$. In the experiments, we employ 3D-printed lattice structures as the solid porous matrix and water as the working fluid. Quantitative analyses of the porous medium Nusselt number $Nu_m$ and local temperature statistics reveal that the present system undergoes a transition through five distinct regimes: I. Conduction, II. Convection, III. Oscillation, IV. Transition, V. Classical Rayleigh--B\'enard convection. This transitional process bridges the gap between Rayleigh--Darcy-like behaviour and Rayleigh--B\'enard-like behaviour in porous media convection. By varying the permeability of the matrix, we further examine the role of the Darcy number $Da$, which turns out to have a profound impact on the transitional processes across different regimes. \textcolor{black}{Flow field measurements reveal that the flow structures within Regime IV and Regime V evolve from several horizontally stacked convection rolls to a single-roll structure, and the pore-scale Reynolds number both exceeds unity in these two regimes.} Finally, we report the corresponding phase diagram in the $Ra$-$Da$ space. 
\end{abstract}



\section{Introduction}
\label{sec:headings}
Convective flow in fluid-saturated porous media constitutes a fundamental process with critical applications spanning geothermal energy extraction, geological sequestration \citep{Huppert2014, Meybodi2015}, subsurface hydrology \citep{Nield2017}, and engineering systems such as packed-bed reactors \citep{Zanganeh2012}. These flows are governed by viscous drag from the solid matrix, resulting in flow dynamics distinct from those of free-fluid flows. A canonical configuration for studying porous media convection is known as Rayleigh--Darcy convection (RDC) system \citep{Nield2017, DePaoli2023}, in which a porous layer saturated with fluid is heated from below and cooled from above. This system also serves as a porous media analogue to Rayleigh--B\'enard convection (RBC), which considers the buoyancy-driven flow in a layer of pure fluid \citep{Ahlers2009, Lohse2010, XIA2013TAML, Xia2024}. Although both systems are driven by buoyancy, they are distinguished by their dominant dissipation mechanisms: in RDC it is the Darcy drag from the solid matrix, whereas in RBC it is the viscous dissipation of the fluid. In addition, the RDC is usually featured by a very small Darcy number, which can be defined as $Da=K/H^2\ll1$, where $K$ is the permeability of the porous medium and $H$ is the height of the system. In fact, RBC can be treated as an extreme case with $Da \rightarrow 1$. However, the inertial effect, which is negligible in RDC, serves as a pivotal factor governing flow dynamics and turbulent fluctuations in RBC.

A central issue in studying various convective flows is determining the quantitative relationship between heat transport efficiency and thermal-driven strength. In RBC, they can be quantified by the fluid Rayleigh number $Ra_f$ and fluid Nusselt number $Nu_f$ as:
\begin{subequations}\label{eq_fluid}
\begin{equation} 
    Ra_f=\frac{\alpha g\Delta TH^3}{\nu_f\kappa_f},
\end{equation}
and
\begin{equation}
    Nu_f=\frac{Q}{k_f\Delta T/H},
\end{equation}
\end{subequations}
where $\alpha$ is the isobaric thermal expansion coefficient of the fluid, $g$ is the gravitational acceleration, $\Delta T$ is the temperature difference between the top and bottom boundaries, $\nu_f$ is the fluid's kinematic viscosity, $\kappa_f$ is the fluid's thermal diffusivity, $Q$ is the total heat flux and $k_f$ is the thermal conductivity of the fluid. Whereas for RDC, the corresponding non-dimensional parameters are the Rayleigh--Darcy number $Ra$ and the medium Nusselt number $Nu_m$: 
\begin{subequations}\label{eq_media}
\begin{equation}
    Ra=\frac{\alpha g\Delta TKH}{\nu_f\kappa_m},
\end{equation}
and
\begin{equation}
    Nu_m=\frac{Q}{k_m\Delta T/H},
\end{equation}
\end{subequations}
where $\kappa_m=(k_m/k_f)\kappa_f$ is the effective thermal diffusivity of the medium and $k_m$ represents the effective thermal conductivity of the medium. \textcolor{black}{Here we have followed the convention in Rayleigh--B\'enard community to use $\kappa_m$ and $\kappa_f$ as the thermal diffusivities. In the porous media convection community, the corresponding notions are often $\alpha_m$ and $\alpha_f$, respectively \citep{Nield2017,Schwendener2026}.} \textcolor{black}{It is also worth mentioning that $k_m$ depends on the thermal conductivity of the fluid phase $k_f$, the solid matrix $k_s$, and the specific geometry of the porous medium.} Combining equations \ref{eq_fluid} and \ref{eq_media}, we can obtain a quantitative connection between RDC and RBC:
\begin{equation} \label{relation}
    Ra=Ra_fDa\lambda, \qquad Nu_m=Nu_f\lambda,
\end{equation}
where $\lambda=k_f/k_m$ is the thermal conductivity ratio. 

Existing experimental investigations on porous media convection have predominantly employed convection cells filled with granular material (\textit{e.g.}, packed beads) \citep{Elder1967, Keene2015}. Measurements of heat transport have revealed a complex $Nu_m-Ra$ relationship \citep{Schneider1965,Elder1967,Combarnous1969,Yen1974,Buretta1976,Lister1990,Kladias1991,Keene2015,Bavandla2024}. Before the onset of convection, the heat is transported by conduction, therefore $Nu_m=1$. When $Ra$ exceeds the onset Rayleigh--Darcy number $Ra_{c}=4\pi^2$, derived by linear stability analysis \citep{Lapwood1948}, convection sets in, and the $Nu_m$ increases almost linearly with increasing $Ra$ as $Nu_m\sim Ra$. The $Nu_m$-$Ra$ curve then exhibits a gradual decrease in its local scaling exponent $\beta = d(\log Nu_m)/d(\log Ra)$ to a value of $\beta\simeq 1/2$. A second transition occurs at $Ra_{c2}$, where the exponent jumps abruptly to a value greater than unity, and the temperature field begins to oscillate \citep{Caltagirone1971}. Subsequent to this jump, $\beta$ undergoes a further transition at $Ra_{c3}$, beyond which it decreases once more and finally converges to an asymptotic scaling exponent. Experiments at very high $Ra$ (up to $2.03\times10^6$) \citep{Keene2015,Ataei2019} have reported asymptotic exponents close to $\beta = 0.3$, which is consistent with the scaling observed in turbulent RBC over a comparable $Ra_f$ range. It is noteworthy that the values of both $Ra_{c,2}$ and $Ra_{c,3}$ exhibit significant scatter across different experiments \citep{Combarnous1969,Buretta1976,Lister1990,Kladias1991}, indicating a strong dependence of the transitions on specific experimental configurations or the Darcy number $Da$. \textcolor{black}{Although the Darcy number $Da$ is known to have a significant influence on porous media convection, the fact that both $Da$ and $Ra$ depend simultaneously on the detailed configuration of the porous medium (e.g., pore size, porosity, porous geometry) poses a great challenge to fully decouple their individual effects. As a result, existing experiments such as packed‑beads systems can hardly clarify the independent influence of $Da$ on heat transport behavior \citep{Bavandla2024,Alam2026}.} Other experimental studies employed the Hele-Shaw cell as an analogue for porous media \citep{Elder1965,Hartline1977} instead, which enables flow visualisation. However, the severe heat loss in such a quasi-two-dimensional system poses a major challenge for accurate heat transport measurements.  

On the other hand, starting from the 1970s \citep{Holst1972}, numerical studies solving the volume-averaged Darcy--Oberbeck--Boussinesq (DOB) equations not only reproduce the first two transitions in the $Nu_m$-$Ra$ relationship as observed in experiments over the low-to-moderate $Ra$ range ($10\le Ra<1000$), but also unravel the underlying flow dynamics. With increasing $Ra$, the system undergoes a sequence of bifurcations, transitioning from steady convection to time-dependent flows, including periodic, quasi-periodic, and chaotic regimes \citep{Caltagirone1981,Caltagirone1989,Graham1994}. The unsteady flow was found to be triggered by the instability of the thermal boundary layers \citep{Graham1994,Otero2004}, leading to a substantial increase in the local scaling exponent $\beta$. In particular, the values of the corresponding transition Rayleigh--Darcy number $Ra_{c2}$ vary across different numerical configurations (2D or 3D) and initial conditions, indicating a more complex underlying mechanism governing this transition.

However, a notable discrepancy arises in the $Nu_m$-$Ra$ relationship between DOB simulations and experimental measurements at high $Ra$ ($Ra > 1000$). Classical theoretical models based on marginal stability \citep{Malkus1954, Howard1966} and rigorous upper-bound analyses \citep{Doering1998, Otero2004} predict an asymptotical scaling of $Nu_m\sim Ra$ at sufficiently high $Ra$. High-$Ra$ DOB simulations (up to $8\times10^4$ for 3-D and \textcolor{black}{$2\times10^5$ for 2-D}) generally confirms this linear scaling behavior \citep{Hewitt2012,Hewitt2014,Pirozzoli2021,DePaoli_CPC_2025}. In contrast, experimental investigations at high $Ra$ have reported an asymptotic scaling exponent of $\beta\approx0.3$---consistent with the turbulent RBC scaling reported earlier. This \textcolor{black}{simulation-experiment} discrepancy \textcolor{black}{originates from the breakdown of Darcy's law, which is the key assumption of DOB models. For convection in porous media, Darcy-type flow is expected when $RaDa^{1/2}\ll1$ and $RaDa^{1/2}/Pr\ll1$ \citep{DePaoli2023}. Once the $Ra$ is sufficiently high, the DOB approximation can no longer accurately describe the flow. Then it must be modified or replaced by other numerical approaches account for pore-scale effects, such as inertial corrections, the influence of `pore-scale parameters' like pore size $l_p$ and porosity $\varphi$ (defined as the volume fraction of the void space in the porous media).} 


More recently, pore-scale direct numerical simulations (DNS) have been used to investigate these pore-scale effects \citep{Karani2017,Liu2020,Korba2022,Xu2023,Schwendener2026}. The simulations are conducted in convective domains containing solid obstacles, which serve as porous media. \citet{Karani2017} compared the pore-scale DNS results for $k_m = k_f = k_s$ (i.e., $\lambda = 1$) with the DOB results, finding an excellent agreement over the range $20 < Ra < 150$. At higher $Ra$, \citet{Liu2020} identified a transition where the scaling exponent $\beta$ changes from near 0.65 to 0.3. They proposed that this transition occurs when the thermal boundary layer thickness $\delta_{th}$ is comparable to the pore size $l_p$, i.e., $\delta_{th}\approx l_p$. \citet{Korba2022} reported a similar scaling transition behaviour, and found that the effective scaling exponent increases as the pore size decreases. \citet{Schwendener2026} reported a three-stage transition of $\beta$ over $100\leq Ra\leq5\times10^4$: from unity to approximately one half, and finally to nearly one third. Nevertheless, the high computational cost of pore-scale DNS restricts these investigations to a limited parameter range ($Da>10^{-6}$) and often necessitates 2D simplifications. These computational limitations in pore-scale DNS also restrain the ability to capture the physical picture of evolution in porous media convection. Conventional experiments on porous media convection, such as those employing packed beads, also face inherent challenges. In practice, accessing a wide range of $Ra$ ($Ra_f$) for a fixed cell height typically requires varying the bead size, which inevitably changes $Da$. This coupling between $Ra$ and $Da$ hinders independent evaluation of the role of Darcy number. Furthermore, the low-$Da$ regime ($Da<10^{-6}$)---critical for natural and industrial porous media applications---has remained largely inaccessible in high-$Ra$ experiments, as it would require both large cell heights and extremely small bead sizes. Finally, the opaque nature of the granular media used in traditional experiments severely hinders detailed flow visualisation, thereby impeding a deeper understanding of the underlying flow transition mechanisms.


\color{black}
Although previous studies have identified the existence of multiple flow regimes in porous media convection with increasing $Ra$, a systematic exploration of their transitions---particularly regarding the influence of the Darcy number---remains incomplete. Both pore-scale simulations and experimental measurements reported that $Da$ has a negligible influence on the onset Rayleigh--Darcy number $Ra_c$ for $\mathcal{O}(10^{-8})\le Da\le \mathcal{O}(10^{-5})$ \citep{Karani2017,Bavandla2024,Bavandla2025}. In contrast, \citet{Alam2026} reported an early onset at $Ra_c\approx20$ for $Da=4.8\times10^{-8}$. They attributed this behaviour to the enhanced local thermal gradients in the fine-grained porous media. Following the onset of convection, existing experimental and numerical data consistently showed that the transitional Rayleigh--Darcy numbers across different regimes generally increase with decreasing $Da$ \citep{Liu2020,Korba2022,Schwendener2026,Combarnous1975,Kladias1991,Bavandla2024,Alam2026}. 

\color{black}

Our study aims to address these unresolved issues. We conduct a series of experiments on porous media convection with an unprecedentedly wide Rayleigh--Darcy number range of $26.8\leq Ra\leq 2.62\times10^5$. To achieve this, we employ 3D-printed periodic lattice structures as the porous matrix, which enables precise and independent control over the medium's geometrical parameters---and thus the Darcy number $Da$. This approach, combined with convection cells of varying heights, allows us to systematically explore the heat transfer and flow dynamics across a broad range of parameters. Furthermore, the current experimental setup also facilitates both flow visualisation and local temperature measurements.

The remainder of this paper is organized as follows: In \S \ref{sec:Set-up}, we describe the experimental setup; then we present the main results in \S \ref{Sec:Result}, which is further divided into four subsections: \S \ref{Sec:Res_Nu} shows the results of heat transport measurements, the identification of distinct flow regimes and the flow states inferred from the local temperature measurements at cell centre are detailed in \S \ref{Sec:Res_Regimes} and \S \ref{Sec:Res_flowstates}, respectively. \textcolor{black}{\S \ref{Sec:Res_Da} presents the effects of the Darcy number.} The flow structures observed within the porous medium are presented in \S \ref{Sec:Res_PIV}. Finally, we summarise and conclude the present study in \S \ref{Sec:Conc}.

\section{Experimental setups}
\label{sec:Set-up}
\begin{figure}
    \centering
    \includegraphics[width=\linewidth]{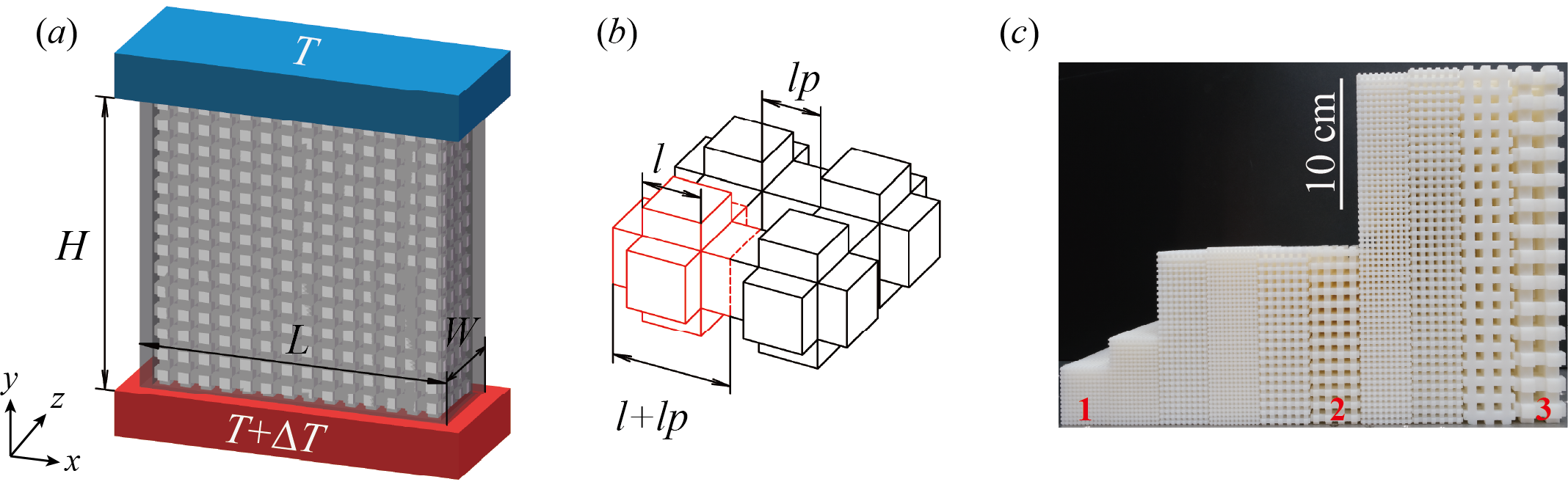}
    \caption{Experimental setups. ({\it a}) Schematic drawing of the convection cell. ({\it b}) A magnified portion of the porous structure. The elemental unit is outlined by the red lines. ({\it c}) A photo of the 3D printed porous samples used in the experiment. Samples labeled with $1$, $2$, and $3$ correspond to the same $Da$ number.}
    \label{fig1}
\end{figure}

In this study, we conducted a series of experiments using a rectangular RBC cell with a fixed cross-section of $L\times W=126\ \text{mm}\times36\ \text{mm}$. The height of the convection cell $H$ varies from 42 mm to 252 mm, so the length-to-height aspect ratio  $\Gamma_{xy}=L/H$ changes from 3 to 1/2 (figure~\ref{fig1}{\it a}). To mimic porous media conditions, we inserted a custom-manufactured porous structure into the convection cell \citep{Zhang2023a, Zhang2023b}. The design of the porous structure, as shown in figure~\ref{fig1}{\it b}, is based on a simple cubic unit cell. The key geometric parameters are the side length $l$ and the pore size $l_p$ of the lattice. The porosity of such a structure can be expressed as, $\varphi=(l^3+3l^2l_p)/(l+l_p)^3$. These matrices were manufactured using stereolithography (SLA) technology with photosensitive resin (Somos LEDO 6060). The density ($\rho_{s}$), thermal conductivity ($k_s$), and specific heat capacity ($c_{P,s}$) of the solid matrix are $\rho_s = 1.13\ \rm{g/cm^3}$, $k_s = 0.2\ \rm{W/(m\cdot K)}$ and $c_{P,s} = 1.5\ \rm{kJ/(kg\cdot K)}$, respectively. The SLA technique enables precise control of $l_p$ and $\varphi$, allowing a systematic exploration of the Darcy number. To determine the permeability $K$ of the porous media, we conducted a series of simulations with different geometric parameters using COMSOL 6.0 and obtained an empirical formula $K=(K_0+Ae^{-\varphi/\varphi_0})l_p^2$ with $K_0=-3.55\times10^{-3}$, $A=3.86\times10^{-3}$, $\varphi_0=-0.41$. Details of these simulations are provided in the Appendix. In this study, the ratio between $l$ and $l_p$ is fixed at unity, yielding a constant porosity of $\varphi=0.5$.

We used deionised water as the working fluid, and maintained the mean temperature at 35.0$^\circ$C, corresponding to a fluid thermal conductivity of $k_f=0.595\ \rm{W/(m\cdot K)}$ and a fluid Prandtl number of $Pr_f=\nu_f/\kappa_f=4.84$ where $\kappa_f$ denotes the thermal diffusivity of the fluid. \textcolor{black}{The thermal conductivity ratio of the porous system is $\lambda = 1.75$, which was obtained from COMSOL simulations (see appendix \ref{appA}). } Both the top and bottom plates of the convection cell are made of copper ($k_{Cu} = 386\ \rm{W/(m\cdot K)}$) with their surfaces coated with nickel to prevent oxidation. The bottom plate is heated by a resistive heater, and the top plate is cooled by a temperature-regulated circulator (PolyScience, Model 9702). Three thermistors (OMEGA, Model 44031) are embedded in each plate to monitor the temperature differences $\Delta T$ across the cell. The sidewall is made of 4 mm thick PMMA ($k_{sw}= 0.21\ \rm{W/(m\cdot K)}$). In addition, Polystyrene foam (approximately 5 cm thick) with a much smaller thermal conductivity around $0.03 \ \rm{W/(m\cdot K)}$ was wrapped around the sidewall to minimise undesired heat leakage to the surroundings. Finally, a copper basin was placed beneath the bottom plate to compensate for heat loss from it. We also placed a small thermistor (TE, Model GAG22K7MCD419) with a head diameter of \SI{300}{\micro\meter} at the cell centre ($x=1/2L$,  $y=1/2H$, $z=1/2W$) to measure the temperature in the fluid $T_c$ at a sampling rate of 5 Hz. \textcolor{black}{For a given Rayleigh--Darcy number, we used the experimental settings of the previous run as the initial condition, and allowed the system to evolve into a statistical steady state prior to data acquisition.}

To explore a wide range of $Ra$ at a fixed Darcy number, we varied the height of the convection cell while keeping the height-to-pore-size ratio $H/l_p$ constant (see table~\ref{table_exp} for details of the geometrical parameters). For comparison, the heat transport of classical RBC was also measured using the same convection cells but without the porous matrix. 

We employed particle image velocimetry (PIV) to quantify the flow structures and velocity fields within the porous medium. A dual Nd:YAG laser (TSI Inc., 532 nm wavelength, 100 mJ per pulse) was used to generate a light sheet with a thickness of about 1 mm. \textcolor{black}{The laser sheet was aligned with the vertical $xy$-plane, and positioned at the mid-width of the cell ($z = 1/2W$), far from both sidewalls. Since the 3D-printed solid porous matrix is opaque, the laser sheet cannot penetrate the lattice structure, so the velocity field measurements are limited to the pore regions within the centre $xy$-plane.} A high-resolution sCMOS camera (Lavision, Imager sCMOS, $2560\times2160$ pixels) was positioned perpendicular to the light sheet to record the particle images. The laser and camera were synchronised by a programmable timing unit (PTU, Lavision, PTU X) controlled via Davis 10 (Lavision) software. The working fluid was seeded with polymaid seeding particles (PSP, Dantex, \SI{50}{\micro m} in diameter, density $1.03-1.05 \ \rm~g/cm^3$) as flow tracers. The Stokes number ranges from $2.1\times10^{-5}$ to $3.3\times10^{-5}$, indicating that the seeding particles can faithfully follow the fluid motion in the present study \citep{Xu2024}. For each experimental run, a sequence of 600 image pairs was recorded at 2 Hz. The velocity field snapshots were obtained by processing the image pairs using a cross-correlation algorithm, and applying a multi-pass processing method with a final interrogation window size of $16 \times 16$ pixels and 50\% overlap, corresponding to a spatial resolution of about 0.5 mm and 1 mm (based on the size of the convection cell).

\begin{table}
  \begin{center}
\def~{\hphantom{0}}
  \begin{tabular}{lccccccccc}
       Case & $l_p$ (mm)   & $l$ (mm) & $\phi$ &$H$ (mm) & $H/l_p$ & $K$ ($\text{m}^2$) & $Da$ & $\it\Gamma_{xy}=L/H$ & $\it\Gamma_{xz}=W/L$\\
       PMC1   & 1.5  & 1.5   & 0.5    & 42 & 28 & $2.13\times 10^{-8}$ & $1.21\times 10^{-5}$ & $3$ & $0.29$\\
       PMC2   & 4.5  & 4.5   & 0.5     & 126 & 28 & $1.92\times 10^{-7}$ & $1.21\times 10^{-5}$ & $1$ & $0.29$\\
       PMC3   & 9    & 9     & 0.5   & 252 & 28 & $7.68\times 10^{-7}$ & $1.21\times 10^{-5}$ & $0.50$ & $0.29$\\
       PMC4   & 1.5  & 1.5   & 0.5     & 63 & 42 & $2.14\times 10^{-8}$ & $5.39\times 10^{-6}$ & $2$ & $0.29$ \\
       PMC5   & 3    & 3     & 0.5  & 126 & 42 & $8.86\times 10^{-8}$ & $5.39\times 10^{-6}$ & $1$ & $0.29$\\
       PMC6   & 6    & 6     & 0.5  & 252 & 42 & $3.42\times 10^{-7}$ & $5.39\times 10^{-6}$ & $0.50$ & 0.29 \\
       PMC7   & 2    & 2     & 0.5   & 124 & 62 & $3.81\times 10^{-8}$ & $2.48\times 10^{-6}$ & $1.02$ & $0.29$ \\
       PMC8   & 1.5  & 1.5   & 0.5       & 126 & 84 & $2.14\times 10^{-8}$ & $1.35\times 10^{-6}$ & $1$ & $0.29$\\
       PMC9   & 3   & 3     & 0.5    & 252 & 84 & $8.57\times 10^{-8}$ & $1.35\times 10^{-6}$ & $2$ & $0.29$ \\
       PMC10   & 2   & 2     & 0.5    & 248 & 124 & $3.80\times 10^{-8}$ & $6.18\times 10^{-7}$ & $0.51$ & 0.29 \\
        \textcolor{black}{RBC1}   & -   & -     & \textcolor{black}{1}    & \textcolor{black}{42} & - & - & - & \textcolor{black}{3} & \textcolor{black}{0.30} \\
       \textcolor{black}{RBC2}   & -   & -     & \textcolor{black}{1}    & \textcolor{black}{64} & - & - & - & \textcolor{black}{1.97} & \textcolor{black}{0.30} \\
       \textcolor{black}{RBC3}   & -   & -     & \textcolor{black}{1}    & \textcolor{black}{126} & - & - & - & \textcolor{black}{1} & \textcolor{black}{0.30} \\
       \textcolor{black}{RBC4}   & -   & -     & \textcolor{black}{1}    & \textcolor{black}{252} & - & - & - & \textcolor{black}{0.50} & \textcolor{black}{0.30} \\
       
  \end{tabular}
  \caption{Details of the 3D-printed structures and convection cells used in experiments.}
  \label{table_exp} 
  \end{center}
\end{table}

\section{Results}
\label{Sec:Result}

\subsection{The global heat transport at $Da=1.21\times10^{-5}$\label{Sec:Res_Nu}}

\begin{figure}
    \centering
    \includegraphics[width=1\linewidth]{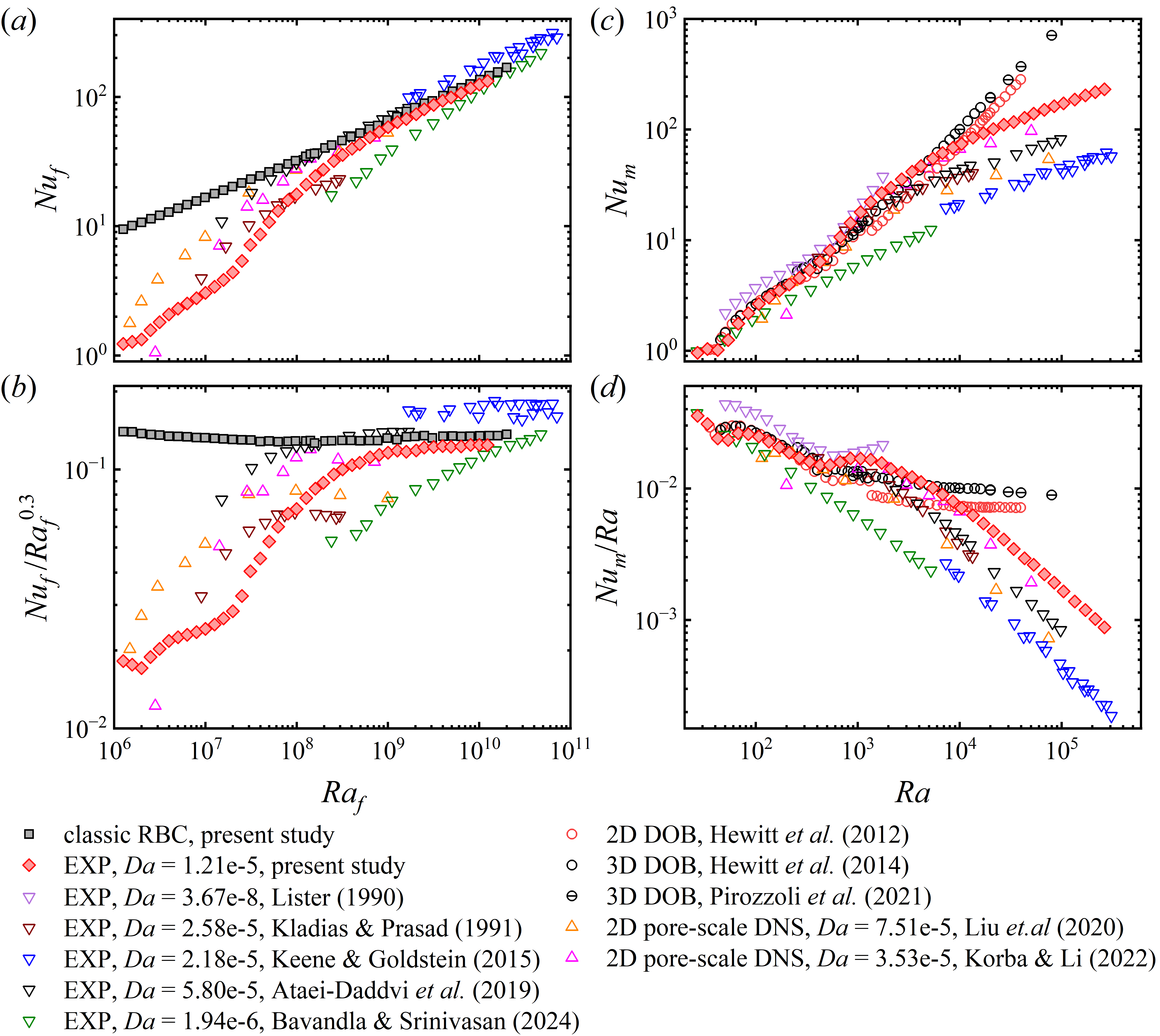}
    \caption{Results of the global heat transport. ({\it a}) Fluid Nusselt number $Nu_f$ versus fluid Rayleigh number $Ra_f$.  ({\it b}) Compensated plots of $Nu_f/Ra_f^{0.3}$ versus $Ra_f$.  ({\it c}) Medium Nusselt number $Nu_m$ versus Rayleigh--Darcy number $Ra$. ({\it d}) Compensated plot of $Nu_m/Ra$ versus $Ra$.}
    \label{fig2}
\end{figure}

We first present the experimentally measured global heat transport data at a fixed Darcy number of $Da = 1.21 \times 10^{-5}$ using red diamonds in figure~\ref {fig2}. The data span over four decades of Rayleigh--Darcy number, enabling us to capture the continuous and complex evolution of heat transport behaviour in porous media convection---from the conduction regime to the RBC-like regime. For ease of reference, literature data (including laboratory experiments, DOB simulations and pore-scale DNS) are also plotted with open symbols. 

Figure~\ref{fig2}({\it a}) shows the relationships between the fluid Nusselt number $Nu_f$ and the fluid Rayleigh numbers $Ra_f$, together with data measured in classical RBC (black squares) for comparison. Figure~\ref{fig2}({\it b}) shows the corresponding compensated plots of $Nu_f/Ra_f^{0.3}$. \textcolor{black}{Firstly, we note that the global heat transfer data obtained from cells with different aspect ratios collapse excellently onto a single master curve (for both PMC and RBC, see table~\ref{table_exp}). This indicates that within the parameter range explored by the present study ($\Gamma_{xz}\ge0.29$), the geometrical confinement effect has a negligible influence on the global heat transport, which is consistent with classical RBC \citep{Huang2013PRL, Xia2023NSR}.}  In particular, our data, previous experimental results, and pore-scale DNS datasets all show a tendency to converge to a regime that features a single power law for sufficiently large $Ra_f$. Specifically, the scaling exponent approaches $\beta\sim0.3$, which is consistent with that of classical RBC (see the solid black symbols).

Figures~\ref{fig2}({\it c}) and ({\it d}) plot the same datasets in terms of the medium Nusselt number $Nu_m$ and the Rayleigh--Darcy number $Ra$. Our experimental data clearly resolve the conductive regime and the onset of convection, with the value of onset Rayleigh--Darcy number agreeing well with the theoretical prediction of $Ra_c=4\pi^2$ \citep{Lapwood1948}. In the moderate-$Ra$ range ($4\pi^2\leq Ra<1000$), our results exhibit remarkable consistency with the DOB data, indicating that the experimental system can be effectively described by the DOB equations. Whereas for $Ra>1000$, our data undergo a gradual transition to the classical RBC scaling of $Nu_m\sim Ra^{0.3}$, deviating significantly from those obtained by DOB simulations, which are close to $Nu_m\sim Ra$. This deviation signifies the breakdown of Darcy's law approximation in the experiments. With increasing $Ra$, the inertial effect---neglected in the DOB equations---becomes more and more important, eventually driving the system toward the classical RBC scaling. The transition from an RDC-like regime to an RBC-like regime also implies a drastic change in the flow structures and the dominant mechanism of energy dissipation. 

\textcolor{black}{In addition to Darcy's law, another key assumption of DOB simulation is local thermal equilibrium (LTE) between the fluid and solid phases. In our experiments, the validity of LTE also depends on the Rayleigh--Darcy number: at small $Ra$, the flow is weak, and the heat exchange between the fluid and solid phases is sufficiently fast to satisfy LTE. As $Ra$ increases, the flow becomes more intense, and the thermal relaxation between the two phases becomes less effective, so that LTE is expected to gradually break down at sufficiently high $Ra$. The progressive departure from LTE may also contribute to the observed discrepancy between our experimental results and existing DOB simulations at high $Ra$.}

\begin{figure}
    \centering
    \includegraphics[width=1\linewidth]{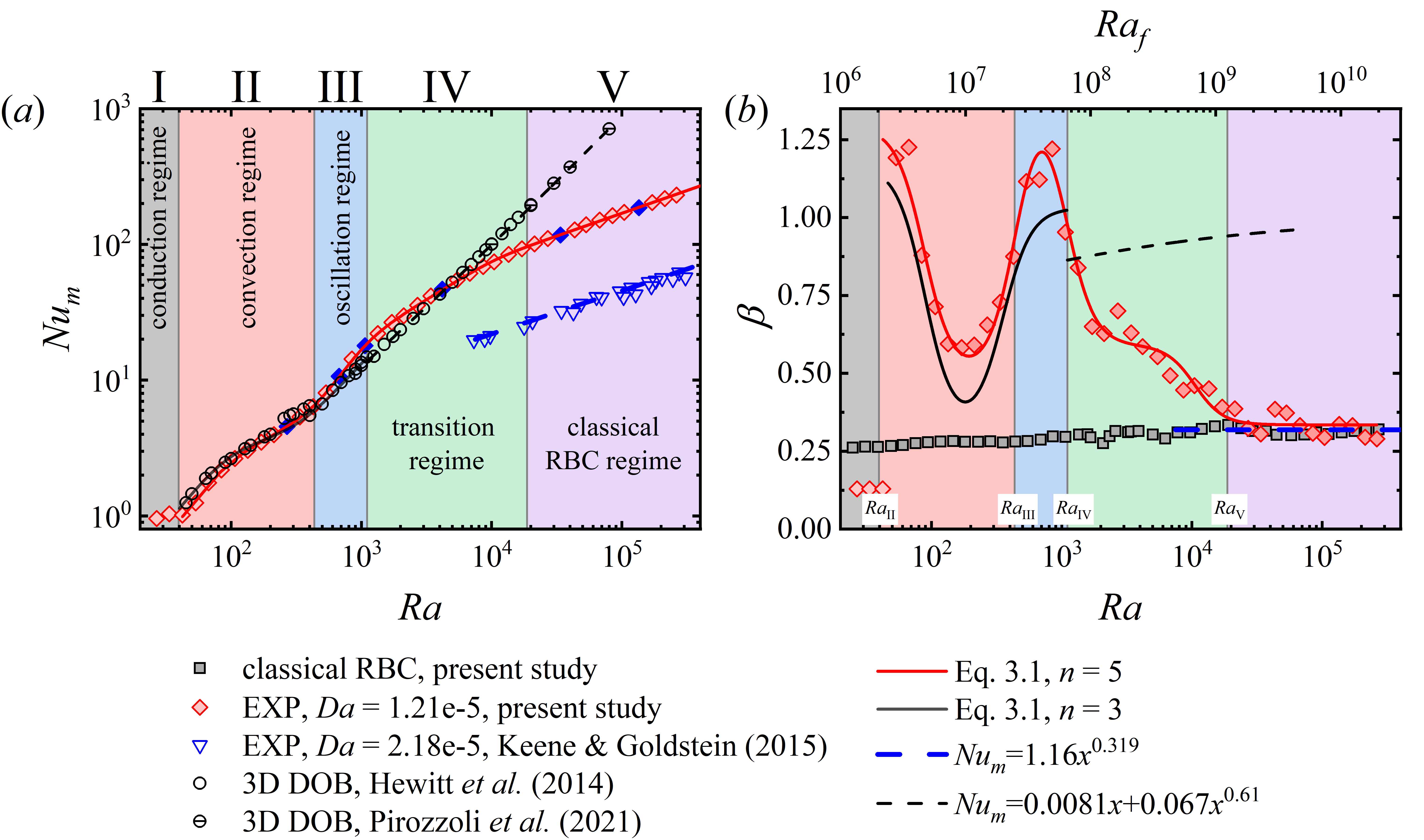}
    \caption{({\it a}) The porous medium Nusselt number $Nu_m$ as a function of Rayleigh--Darcy number $Ra$. The solid lines show fittings with equation~\ref{trans}. The dashed black line and \textcolor{black}{blue} line show the fitting functions adopted from references~\citet{Pirozzoli2021} and~\citet{Keene2015}, respectively. The blue solid diamonds represent values of $Ra$ at which the centre temperature measurements are conducted. ({\it b}) The local scaling exponent $\beta$ of the $Nu_m$-$Ra$ curve as a function of $Ra$. \textcolor{black}{The solid lines correspond to those in panel (a).} The data obtained from classical RBC are plotted with solid squares for benchmarking, \textcolor{black}{where the $Ra_f$ axis is located at the top of the figure.}}
    \label{fig3}
\end{figure}

\subsection{Regimes of heat transport}
\label{Sec:Res_Regimes}
For a quantitative analysis of the transitional behaviour of heat transport, we adopt the following $n$th-order transition function \citep{Zhang2023b}:
\begin{equation}
    \label{trans}
    Nu_m = A\cdot Ra^{\beta_n}\cdot \left[1+\left(\frac{Ra_{t,1}}{Ra}\right)^4\right]^{-\frac{\beta_1-\beta_2}{4}} \cdots \left[1+\left(\frac{Ra_{t,n-1}}{Ra}\right)^4\right]^{-\frac{\beta_{n-1}-\beta_{n}}{4}}, \qquad n\geq2,
\end{equation}
where $A$, $\beta_{1...n}$, $Ra_{t,1 ... n-1}$ are $2n$ free fitting parameters. This function exhibits the following scalings: $Nu_m\sim Ra^{\beta_1}$ for $Ra\ll Ra_{t,1}$; $Nu_m\sim Ra^{\beta_i}$ for intermediate $Ra$ regimes ($Ra_{t,i-1}\ll Ra\ll Ra_{t,i}$); and $Nu_m\sim Ra^{\beta_n}$ for $Ra\gg Ra_{t,n-1}$. Thus, it allows $Nu_m$ to follow different power-law scalings (with exponent $\beta_1$ to $\beta_n$) across successive $Ra$ regimes.

We fitted our experimental data over the range $4\pi^2<Ra\leq 2.62\times 10^5$ using Eq. \ref{trans} with $n$ up to 5. For comparison, the 3D DOB data of \citet{Hewitt2014}(over $4\pi^2<Ra<1000$) were fitted with a 3rd-order form of equation \ref{trans}. The fitting curves are plotted as solid lines in figure~\ref{fig3}({\it a}) and the corresponding fitting parameters are listed in table~\ref{trans_com}.

\begin{table}
  \begin{center}
\def~{\hphantom{0}}
  \begin{tabular}{ccccccccccc}
     & $A$ & $Ra_{t,1}$& $Ra_{t,2}$ & $Ra_{t,3}$& $Ra_{t,4}$ & $\beta_1$& $\beta_2$& $\beta_3$& $\beta_4$& $\beta_5$ \\
    \makecell{Present study \\ ($Da=1.21\times10^{-5}$)} & 3.59 & 87.7 & 455 & $1.05\times10^3$ & $1.05\times10^4$ & 1.29 & 0.491 & 1.522 & 0.589 & 0.335 \\
    \makecell{3-D DOB \\ \citep{Hewitt2014}} & 0.01 & 89.3 & 351 & - & - & 1.17 & 0.313 & 1.03 & - & - \\
  \end{tabular}
  \caption{Fitting parameters of the transition functions.}
  \label{trans_com}
  \end{center}
\end{table}

As shown in figure~\ref{fig3}({\it a}), the 5th-order transition function provides a precise quantitative description of the present experimental data across the entire $Ra$ range, \textcolor{black}{validating its suitability for describing the multi-stage transitional heat transport in porous media convection.} For a clearer view of the transition, we plot in figure~\ref{fig3}({\it b}) the corresponding local scaling exponent $\beta$. The solid curves are extracted from the derivatives of the fitted equations, while the diamonds are determined using a power-law fitting to a sliding window of three adjacent experimental data points. The excellent agreement between these two approaches justifies the rationality of selecting a fitting function with $n=5$. 

With the above quantitative description of $\beta$-$Ra$ relationship, we can identify five distinct regimes, marked by different colours in figures~\ref{fig3}({\it a}) and ({\it b}). The critical Rayleigh--Darcy numbers demarcating these regimes are denoted as $Ra_{\rm II}$, $Ra_{\rm III}$, $Ra_{\rm IV}$ and $Ra_{\rm V}$. The first critical point $Ra_{\rm II}$ marks the onset of convection and is consistent with the theoretical prediction of $4\pi^2$. \textcolor{black}{The second and third critical points, $Ra_{\rm III}$ and $Ra_{\rm IV}$ were determined from the local extrema of $d\beta/d(\log Ra)$. To determine $Ra_{\rm V}$, we extended the tangent to $\beta(Ra)$ at the last minimum of $d\beta/d(\log Ra)$ until it intersected the asymptotic value $\beta=\beta_5$, and defined this intersection as $Ra_{\rm V}$. The details are provided in appendix B.} It is also worth mentioning that the values $Ra_{\rm III}$, $Ra_{\rm IV}$ and $Ra_{\rm V}$ are not identical to the fitting parameters of equation~\ref{trans}.

For $0\leq Ra\leq Ra_{\rm II}=4\pi^2$, heat transport is purely conductive, with $Nu_m=1$ and $\beta\approx0$. Convection sets in at $Ra_{\rm II}$. The exponent $\beta$ jumps sharply to a value greater than unity and then gradually decreases to approximately 0.5 as $Ra$ increases to around $Ra_{\rm III}$. As $Ra$ exceeds $Ra_{\rm III}$, $\beta$ exhibits a significant jump to a value greater than one. Beyond $Ra_{\rm IV}$, the local scaling exponent again shows a mild decrease. Finally, for $Ra$ greater than $Ra_{\text{V}}$, $\beta$ approaches to a value of approximately 0.335.

Among the identified regimes, Regime I corresponds to the conductive state before the onset of convection. The characteristics of Regime II and Regime III are consistent with the results obtained in RDC, as evidenced by the close agreement between the experimental and DOB $\beta$-$Ra$ curves in the corresponding $Ra$ range (figure~\ref{fig3}({\it b})). \textcolor{black}{This indicates that the system's behaviour in these two regimes is primarily governed by Darcy's law, and also justifies them as the convection regime (II) and oscillation regime (III) (we will confirm this point in \S\ref{Sec:Res_flowstates})}. In contrast, the asymptotic scaling in Regime V closely aligns with the heat transport scaling of the classical RBC system, indicating that the system's dynamics are now dominated by inertial effects and viscous dissipation. Therefore, Regime V is identified as the classical RBC regime. Lastly, Regime IV represents the transitional process where the scaling exponent $\beta$ decreases systematically from RDC-like scaling toward the RBC-like asymptotic value. Physically, this regime corresponds to the system's transition from Darcy-drag-dominated to inertia-dominated dynamics. We therefore identify it as the transition regime, which serves as a bridge connecting the RDC and RBC systems.

To summarise, the application of the transition function (equation \ref{trans}) has enabled a quantitative identification of five consecutive regimes: I. conduction regime; II. convection regime; III. oscillation regime; IV. transition regime; and V. classical RBC regime.

\textcolor{black}{It is noteworthy that the confinement parameter $\Lambda$, proposed by \citet{Noto_PNAS_2024} and first introduced to porous media convection by \citet{Schwendener2026}, provides an alternative perspective for interpreting the flow transition. Defining $\Lambda=\delta_{th}/l_p=(H/l_p)/(2Nu_m)$ as the ratio of the thermal boundary-layer thickness to the pore size, we find that the critical Rayleigh--Darcy number $Ra_{\rm IV}$ corresponds to $\Lambda \approx 1.5$, i.e. $\Lambda=O(1)$. This suggests that the transition from Regime III to Regime IV occurs when the thermal boundary layer becomes comparable to the pore scale, and can thus be interpreted as a consequence of plume-scale confinement relaxation.}

\subsection{Flow states at different regimes}
\label{Sec:Res_flowstates}
\begin{figure}
    \centering
    \includegraphics[width=0.8\linewidth]{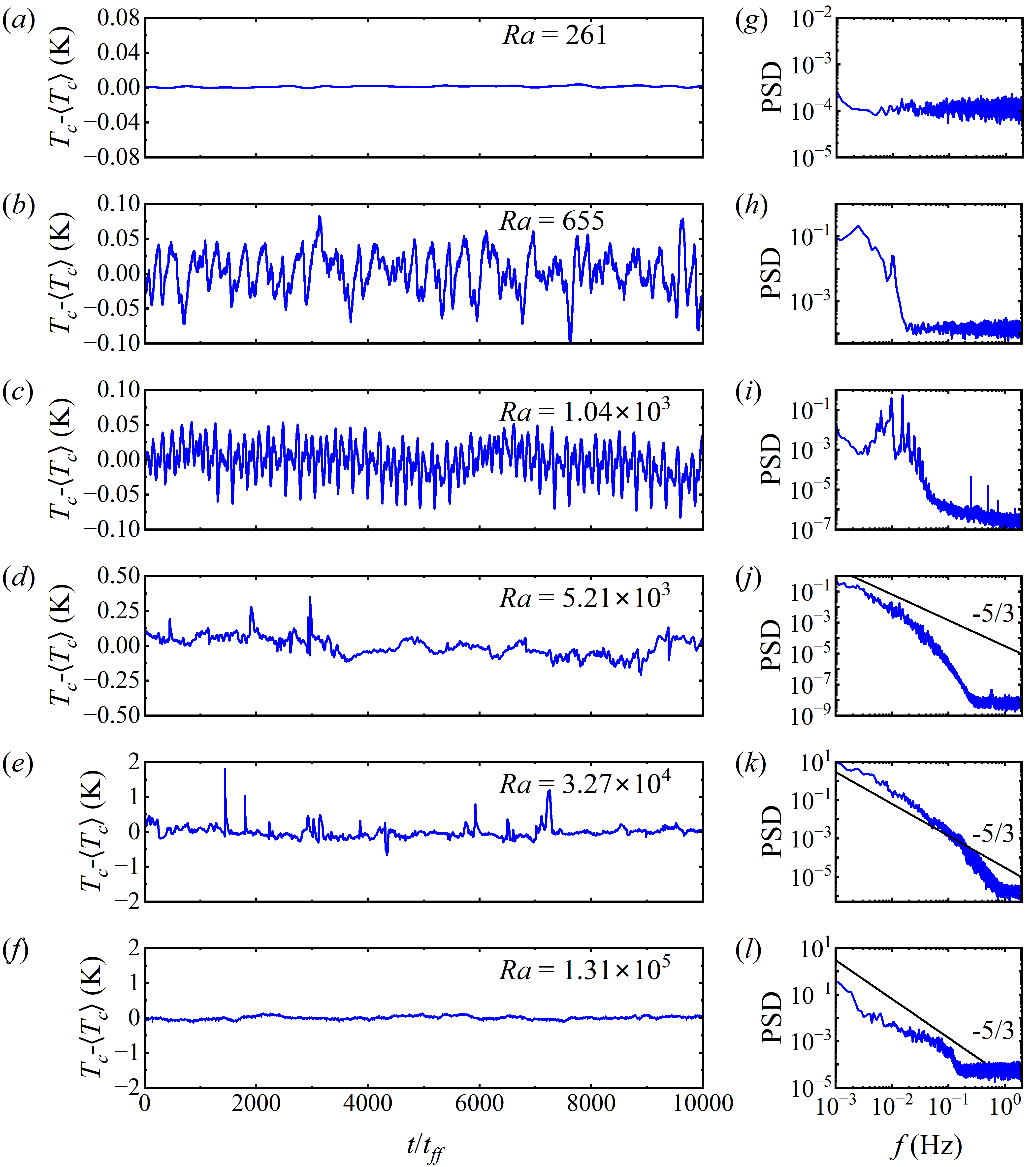}
    \caption{({\it a-f}) The time series of the experimentally measured temperature fluctuation $T_c-\left\langle T_c\right\rangle$ at the cell centre for different values of $Ra$. The horizontal axis are normalised by the free-fall time $t_{ff}=\sqrt{H/(\alpha g\Delta T)}$. ({\it g-l}) The PSDs of the corresponding time series in ({\it a-f}). The black line denotes the $f^{-5/3}$ power law.}
    \label{fig4}
\end{figure}

In this subsection, we present results from centre-temperature ($T_c$) measurements, which shed light on the evolution of the flow state with increasing $Ra$.

Figure~\ref{fig4}({\it a-f}) shows the time series of $T_c-\left\langle T_c\right\rangle$ measured at six representative values of $Ra$---marked as blue solid diamonds in figure~\ref{fig3}({\it a}). At $Ra=261$ (within Regime II), the centre temperature time series (figure~\ref{fig4}({\it a})) remains nearly constant, with its corresponding power spectrum distribution (PSD) exhibiting white noise characteristics (figure~\ref{fig4}({\it g})). This time-independent local temperature behaviour confirms that the flow is in a steady convection state. As $Ra$ increases to $Ra=655$, the system enters the oscillation regime (Regime III). The centre temperature displays low-frequency, small-amplitude oscillations (figure~\ref{fig4}({\it b})), which is also reflected in its PSD (figure~\ref{fig4}({\it h})). A further increase of $Ra$ to $Ra=1.04\times10^3$ (still within Regime III) excites oscillations with multiple harmonic frequencies (see figures~\ref{fig4}({\it c} and {\it i})). 

When the system is in the transition regime ($Ra=5.21\times10^3$), the time series of centre temperature shows irregular, intermittent fluctuations (see figure~\ref{fig4}({\it d})). The corresponding PSD (figure~\ref{fig4}({\it j})) shows a smooth broadband distribution, indicating a transition toward turbulence. At $Ra=3.27\times10^4$, the time series of $T_c$ displays more intense fluctuations (figure~\ref{fig4}({\it e})), with its PSD (figure~\ref{fig4}({\it k})) exhibiting a power law scaling close to $k^{-5/3}$---a hallmark of turbulence \citep{Sano1989,Niemela2000,Zhou2001}.

Notably, at $Ra=1.31\times10^5$, the intensity of the fluctuation of the centre temperature is significantly lower than that at $Ra=3.27\times10^4$. This is attributed to the fact that measurements for these two $Ra$ were taken in cells with different aspect ratios ($\Gamma_{xy}$) (see table~\ref{table_exp}): The time series in figure~\ref{fig4}({\it e} and {\it k}) were obtained in a cell with $\Gamma_{xy}=1$, while those of figure~\ref{fig4}({\it f} and {\it l}) were measured in a cell with $\Gamma_{xy}=1/2$ instead. In convective flows, domains with different aspect ratios usually develop distinct large-scale flow structures \citep{Xi2008,vanderPoel2012,Wang2020}. Therefore, for these two sets of local temperature measurements, the `geometric centre' may correspond to regions with different large-scale flow characteristics, leading to the seemingly singular temperature fluctuations behaviour. This physical picture is validated by preliminary results of flow visualisation, which will be presented later in \S\ref{Sec:Res_PIV}.

\subsection{Effect of Darcy number $Da$}
\label{Sec:Res_Da}
As mentioned in $\S$ \ref{sec:headings}, one of the central advantages of employing 3D-printed matrices as the porous medium is that it enables precise control of the Darcy number $Da$. In this subsection, we will explore in detail how the Darcy number affects the transitions between different flow regimes. To achieve this, we conducted a series of additional experiments with different combinations of $l_p$ and $H$, covering a Darcy number range of $6.18\times10^{-7}\leq Da \leq 1.21\times10^{-5}$, while maintaining a constant porosity of $\varphi = 0.5$.

\begin{figure}
    \centering
    \includegraphics[width=1\linewidth]{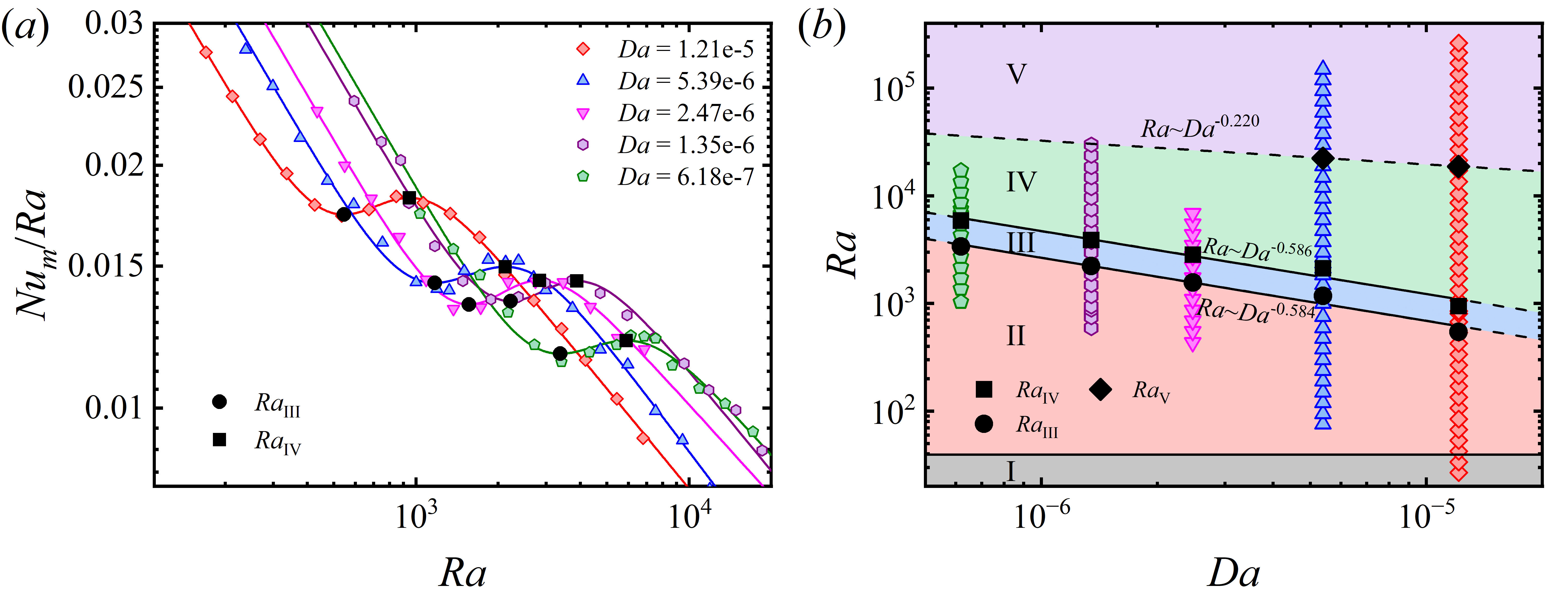}
    \caption{({\it a}) Compensated plot of the experimentally measured media Nusselt number $Nu_m/Ra$ versus $Ra$ for different values of $Da$. The solid lines represent fitting curves of equation~\ref{trans} with $n=3$. ({\it b}) The phase diagram of different heat transport regimes in the $Ra$-$Da$ space. \textcolor{black}{The solid symbols in both panel (a) and panel (b) present the critical Rayleigh--Darcy numbers.}}
    \label{fig5}
\end{figure}

\begin{table}
  \begin{center}
\def~{\hphantom{0}}
  \begin{tabular}{ccccccccc}
    $Da$& $A$& $Ra_{t,1}$& $Ra_{t,2}$& $\beta_1$& $\beta_2$& $\beta_3$& $Ra_{\rm III}$& $Ra_{\rm IV}$\\
    $1.21\times10^{-5}$& 0.353& $5.45\times10^2$& $8.65\times10^2$& 0.417& 1.96& 0.59& $5.50\times10^2$& $8.60\times10^2$\\
    $5.39\times10^{-6}$& 0.553& $1.04\times10^3$& $2.23\times10^3$& 0.417& 1.47& 0.55& $1.05\times10^3$&$ 2.30\times10^3$\\
    $2.47\times10^{-6}$&0.318&$1.90\times10^3$&$2.09\times10^3$&0.417&7.04&0.63&$1.91\times10^3$&$2.10\times10^3$\\
    $1.35\times10^{-6}$&0.501&$2.49\times10^3$&$3.17\times10^3$&0.417&3.29&0.59&$2.50\times10^3$&$3.10\times10^3$\\
    $6.18\times10^{-7}$&0.376&$3.00\times10^3$&$5.95\times10^3$&0.417&1.50&0.62&$3.02\times10^3$&$6.05\times10^3$\\
  \end{tabular}
  \caption{Fitting parameters of the transition functions for different Darcy numbers.}
  \label{trans_Da}
  \end{center}
\end{table}

Figure~\ref{fig5} ({\it a}) presents the compensated medium Nusselt number, $Nu_m/Ra$, as a function of $Ra$ for different $Da$ values. In general, the five datasets exhibit nearly identical shapes but shift toward the lower right as the Darcy number decreases. For quantitative analysis, each dataset (within a Rayleigh--Darcy number range of $100<Ra<2\times10^4$) was fitted using the third-order transition function (equation~\ref{trans}, $n=3$). The fitting results are plotted as solid curves in figure~\ref{fig5}({\it a}). The two critical Rayleigh--Darcy numbers, $Ra_{\rm III}$ and $Ra_{\rm IV}$ \textcolor{black}{(marked by solid circles and squares)}, were determined for each $Da$ using the same method detailed in \S\ref{Sec:Res_Regimes}. \color{black} Specifically, we extracted the local scaling exponent $\beta$ as a function of $Ra$ from the derivative of the 3rd-order transition function and identified the critical $Ra$ numbers by finding the extrema of $d\beta/d(\log Ra)$. This approach enables a direct comparison of the transitional behaviour across different $Da$. \color{black}

For the Darcy number range explored in this study, it is seen that the critical Rayleigh--Darcy numbers ($Ra_{\rm III}$ and $Ra_{\rm IV}$)---marking the transitions from Regime II to Regime III and from Regime III to Regime IV---both shift to higher values with decreasing $Da$ (\textit{i.e.} reduced permeability). Furthermore, both $Ra_{\rm III}$ (onset of time-dependent flow) and $Ra_{\rm IV}$ (onset of the transition to RBC) exhibit power-law dependencies on $Da$ with $Ra_{\rm III}\sim Da^{-0.584}$ and $Ra_{\rm IV}\sim Da^{-0.586}$, respectively (see the black solid lines in figure~\ref{fig5}({\it b}). The exponents are nearly identical, implying a consistent mechanism governing both transitions. 

Previous studies of pore-scale DNS suggest that the transition from Regime III to Regime IV occurs when the thermal boundary layer thickness $\delta_{th}$ becomes comparable to the pore size $l_p$ \citep{Liu2020}, \textcolor{black}{which corresponds to a confinement parameter of order unity $\Lambda = O(1)$}. Building on this hypothesis, if we make a first-order assumption that the $Nu_m$-$Ra$ relationship in both Regime II and Regime III follows a single power-law scaling $Nu_m\sim Ra^{\bar\beta}$, and consider the fact that $Da=K/H^2\sim l_p^2/H^2$ and $\delta_{th}/H=1/2Nu_m$, we can obtain the following relationship for the transitional Rayleigh--Darcy number associated with $\delta_{th}\sim l_p$:

\begin{equation}
    Ra_{\rm IV}\sim Da^{-1/(2\bar\beta)}.
\end{equation}

The experimentally measured scaling exponent $Ra_{\rm IV}\sim Da^{-0.586}$ yields $\bar\beta = 0.853$, which is consistent with the measured local scaling exponents (figure \ref{fig3}({\it b})). This result further validates the hypothesis that the transition from Regime III to Regime IV is associated with the crossover between the thermal boundary layer thickness and the pore size. 

In addition, the critical Rayleigh--Darcy number $Ra_{\rm V}$ for $Da=5.39\times10^{-6}$ was also obtained from its $Nu_m(Ra)$ fitting curve. The boundary between Regimes IV and V in the phase diagram (dashed line in figure~\ref{fig5}({\it b})) was then estimated by connecting the two solid diamonds with a power law. This power-law dependence of $Ra_{\rm V}$ on $Da$, $Ra_{\rm V}\sim Da^{-0.220}$, differs significantly from those of $Ra_{\rm III}$ and $Ra_{\rm IV}$, indicating a distinct physical mechanism governing the transition to the classical RBC regime.

With the above, we can now construct a phase diagram of the flow states in the $Ra$-$Da$ parameter space, as shown in figure~\ref{fig5} ({\it b}). This diagram clarifies the role of the Darcy number: it shows a significant influence on the transition between Regimes II, III, and IV. At low $Da$ (dense porous media), the system exhibits characteristic RDC behaviour (Regimes II and III) over a relatively wide range of $Ra$; in contrast, at high $Da$ (sparse porous media), the system evolves into the transition regime (IV) and the classical RBC regime (V) at a substantially lower $Ra$. Although the critical point that separates Regime III and Regime IV can be rationalised by equalising the thermal boundary layer thickness $\delta_{th}$ and the pore size $l_p$, the physical mechanisms governing the transitions among other regimes remain unresolved and merit future investigation. Nevertheless, our results successfully reconcile the seemingly contradictory findings from DOB simulations, packed-bead experiments, and pore-scale DNS.

Beyond the five flow regimes discussed above, there is an alternative classification of flow regimes based on the relative importance of the inertial effects: the Darcy regime and the Forchheimer regime \citep{Nield2017}. In the former, flow is governed by Darcy's law, with buoyancy force balanced by the Darcy drag, resulting in the linear heat transport scaling, $Nu_m\sim Ra$. As flow intensity increases, the linearity of Darcy's law gradually breaks down, and a modification with Forchheimer drag becomes necessary \citep{Joseph1982}. In the Forchheimer regime, a balance between buoyancy and Forchheimer drag yields $Nu_m\sim Ra^{1/2}Pr_p^{1/2}$, where $Pr_p=Pr_fc_F^{-1}Da^{-1/2}\lambda$ is the modified Prandtl number with $c_F$ being the Forchheimer coefficient \citep{Wang1987}. The transition point from the Darcy regime to the Forchheimer regime can be estimated by equating the two scaling relationships above, yielding $Ra_{D-F}\sim Pr_p$. Substituting the definition of the modified Prandtl number, we can obtain $Ra_{D-F}\sim Da^{-1/2}$, which is \textcolor{black}{generally consistent with the transition from Regime III to Regime IV identified in this study.}

\subsection{The flow structures in Regimes IV and V}
\label{Sec:Res_PIV}

\begin{figure}
    \centering
    \includegraphics[width=0.95\linewidth]{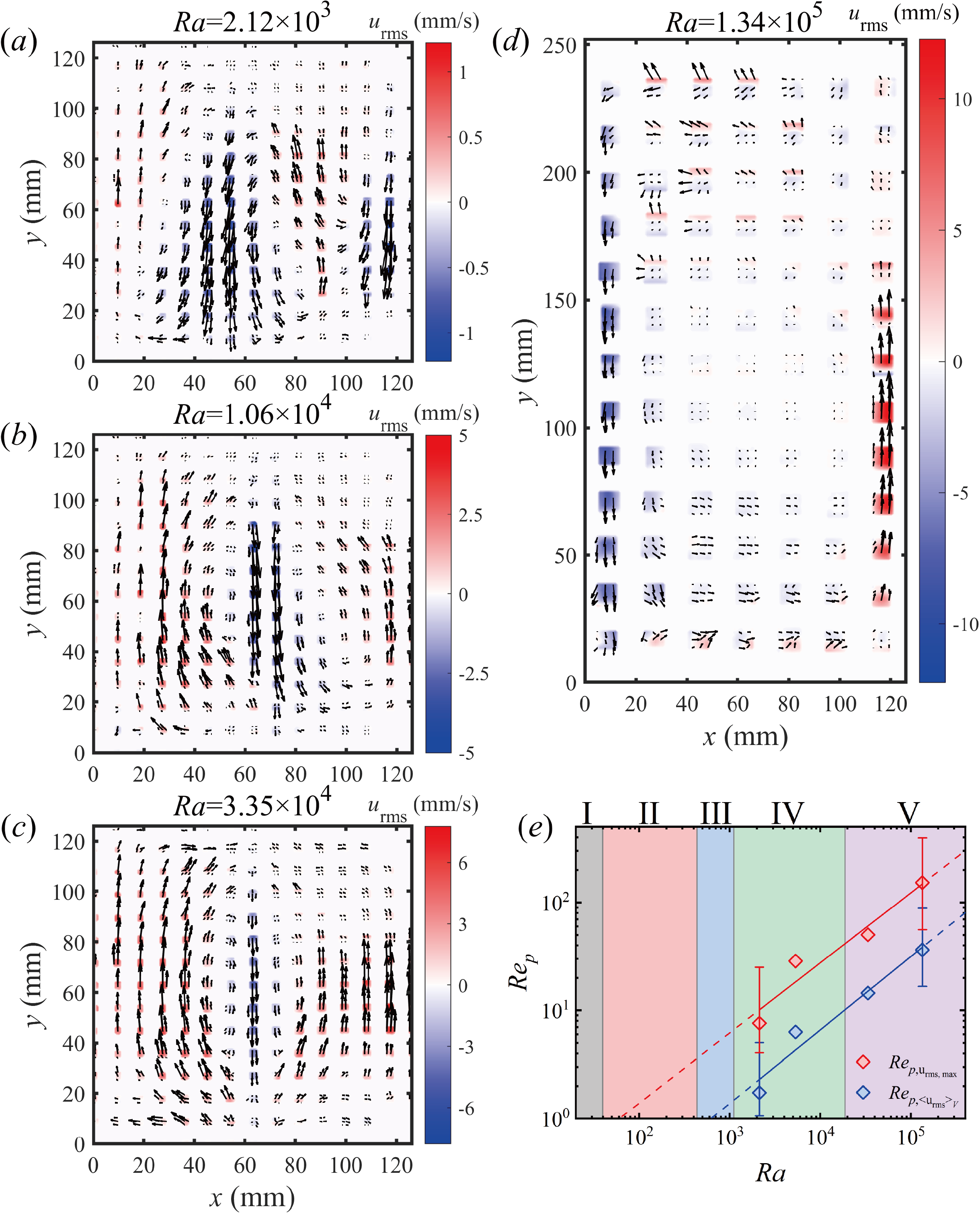}
    \caption{({\it a}-{\it d}) Experimentally measured velocity fields at the centre $xy$-plane ($z=1/2W$) for ({\it a}) $Ra=2.12\times10^3$, ({\it b}) $Ra=1.06\times10^4$ and ({\it c}) $Ra=3.35\times10^4$ in a $\Gamma_{xy}=1$ cell; ({\it d}) $Ra=1.34\times10^5$ in a $\Gamma_{xy}=1/2$ cell. ({\it e}) Pore-scale Reynolds number $Re_p$ based on maximum or volume-averaged root-mean-square velocity as a function of Rayleigh--Darcy number $Ra$. The solid lines represent the power-law fit to the experimental data as a first-order approximation, and the dashed lines show the extrapolation of the fitted function. \textcolor{black}{Error bars at both ends of the fitted curve represent a 90\% confidence interval.} The Darcy number is fixed at $Da=1.21\times10^{-5}$.}
    \label{fig6}
\end{figure}

As mentioned above, another key advantage of the present experimental approach is that it enables visualisation of the flow field within the porous media. In this subsection, we present preliminary results of the global flow structure measured via PIV. The measurements were conducted in the vertical $xy$-plane located at $z=1/2W$ of cell No.2 and cell No.3 (see table~\ref{table_exp}). The investigated values of the Rayleigh--Darcy number $Ra=2.12\times10^3$, \textcolor{black}{$1.06\times10^4$}, $3.35\times10^4$ and $1.34\times10^5$ cover the transition regime and classical RBC regime.  

Figures~\ref{fig6} ({\it a-d}) show the measured velocity fields (averaged over about one minute) within the porous media. The flow is seen to evolve from several horizontally stacked convection rolls to a single large-scale circulation. The number of rolls $n$, with a typical horizontal length scale of $l=L/n$, appears to decrease with increasing $Ra$. On the other hand, the non-dimensional horizontal length scale of the flow structures---$l/H=(L/n)/H=\Gamma_{xy}/n$---exhibits a \textcolor{black}{monotonic non-decreasing trend---it increases from approximately 1/3 (figure~\ref{fig6}({\it a})) to 1/2 (figure~\ref{fig6}({\it b})), and then stays constant as $Ra$ further increase (see figure~\ref{fig6}({\it b-d}))}. This is in contrast to DOB simulations, which report a negative power-law scaling between $l/H$ and $Ra$ \citep{Wen2015,DePaoli2022,Zhu2024}. \textcolor{black}{This discrepancy originates from the fact that DOB simulations rely on the Darcy law (neglecting inertia), whereas the present PIV measurements are conducted in Regime IV and Regime V (see figure~\ref{fig6}({\it e})), where the Darcy law is no longer valid.} Notably, even in the classical RBC regime, the global flow structures \textcolor{black}{obtained by the present study} (see figures~\ref{fig6}({\it c}) and \ref{fig6}({\it d})) are apparently different from those measured in a classical RBC system, where a single large-scale circulation (for cells with unity aspect ratio) \citep{Xia2003,Xi2008PRE} or two vertically stacked convection rolls (for cell with aspect ratio one half) \citep{Stringano2006,Xi2008} are reported.

\textcolor{black}{Furthermore, we calculated the pore-scale Reynolds number $Re_p$ based on the volume-averaged root-mean-square (rms) velocity, $Re_{p,\langle u_{\text{rms}}\rangle_V}=\langle u_{\text{rms}}\rangle_Vl_p/\nu_f$, and that based on the maximum rms velocity $Re_{p,u_{\text{rms,max}}}= u_{\text{rms,max}}l_p/\nu_f$. Figure~\ref{fig6}({\it e}) shows $Re_p$ as a function of the Rayleigh--Darcy number $Ra$. For all explored $Ra$, both definitions of pore-scale Reynolds number are considerably larger than unity, indicating that Darcy's law is invalid---consistent with our heat transport flow regime classification. In addition, both $Re_{p,\langle u_{\rm rms}\rangle_V}$ and $Re_{p,u_{\rm rms,max}}$ exhibit power-law relationships with $Ra$, with best fits yielding $Re_{p,u_{\rm rms,max}}\sim Ra^{0.648}$ and $Re_{p,\langle u_{\rm rms}\rangle_V}\sim Ra^{0.678}$, respectively. \textcolor{black}{It is worth mentioning that $Re_p$ is evaluated based on PIV measurements within the porous structure, where the optical interrogation window is limited by the solid matrix, leading to considerable uncertainties. Therefore, the single power-law fits for $Re_p$-$Ra$ relationships only serve as a first-order approximation. Nevertheless, the obtained scaling exponents} are slightly higher than the 1/2 value predicted for the Forchheimer regime. And the intersection points of the two extrapolated curves with $Re_p=1$ are close to the onset of convection and oscillation, respectively.}


\textcolor{black}{It is also worth mentioning that mechanical dispersion \citep{DePaoli2023}, an inherent effect in porous media convection, may also influence the transitions between different flow regimes. Mechanical dispersion arises from spatial variability in pore-scale velocities and is responsible for the broadening of thermal plumes induced by inter-pore and intra-pore advection. A quantitative characterisation of mechanical dispersion would require independent measurement of the mechanical dispersion coefficient, or introducing an additional dimensionless parameter---the dispersion Rayleigh--Darcy number $Ra_d$ \citep{Liang_GRL_2018}. However, this demands high-resolution flow field and temperature field measurement, and therefore is beyond the scope of the present experimental study.}

\section{Summary and conclusions \label{Sec:Conc}}
\textcolor{black}{This study presents a systematic investigation of heat transport and flow dynamics in porous media convection, combining global heat transport, local temperature statistics, and PIV measurements. We employed custom-designed 3D-printed porous matrices and several convection cells with different heights to precisely and independently vary the pore size $l_p$ and Darcy number $Da$ at a fixed porosity $\varphi=0.5$. This scheme enables us to explore an unprecedentedly wide parameter space: $26.8 \le Ra \le 2.62\times10^5$ and $6.18\times10^{-7} \le Da \le 1.21\times10^{-5}$.}

Our main findings are as follows. First, we quantitatively characterised the global heat transport and the local temperature fluctuations in porous media convection, based on which we identified five distinct regimes: I. conduction regime, II. convection regime, III. oscillation regime, IV. transition regime, and finally V. classical RBC regime. \textcolor{black}{And the transition from Regime III to Regime IV can be interpreted as a result of plume-scale confinement relaxation.} An increase in $Da$ promotes the onset of time-dependent flow and shifts the transition from the RDC-like regime to the RBC-like regime towards significantly lower $Ra$. The critical Rayleigh--Darcy number $Ra_{\rm III}$ for the transition from Regime III to Regime IV shows a power-law scaling of $Ra\sim Da^{-0.586}$, which is generally consistent with the physical picture that such a transition occurs when the thermal boundary layer thickness is comparable to the porous medium's pore scale. We also constructed a $Ra$-$Da$ phase diagram for porous media convection, providing new insights into convective transport and flow characteristics across different permeabilities and varying thermal driving strengths. \textcolor{black}{Lastly, PIV measurements reveal that the flow structures in Regime IV and Regime V are characterised by several horizontally stacked convection rolls, and the number of rolls seems to decrease with increasing $Ra$, and finally merges into a single roll. And the pore-scale Reynolds numbers in Regime IV and Regime V are both larger than unity, demonstrating the importance of inertia in these two regimes.} 

This work successfully bridges the gap between RDC and RBC by experimentally demonstrating how and when the system's heat transfer property transitions from RDC-like to RBC-like. In addition, the excellent agreement between our experimental data, DOB simulations, pore-scale DNS, and prior experimental results also validates our experimental approach, i.e. utilising 3-D printed porous structures, as an effective tool for fundamental research on porous media convection.

\textcolor{black}{Nevertheless, the problem of porous media convection is extremely complex as it involves a lot of physical parameters. And the present study only explored a limited parameter space, mainly focusing on the effects of $Ra$ and $Da$. Recent studies \citep{Bavandla2025,Schwendener2026} have also addressed the importance of porosity and thermal 
conductivity ratio in determining the flow evolution. From this perspective, the approach of using 3D-printed porous matrix also opens several avenues for future study---it provides a flexible experimental platform that can be used to systematically explore other flow parameters and flow physics like mechanical dispersion.} Furthermore, the physical mechanisms underlying inter-regime transitions, especially the destabilization between Regime II and Regime III, and the scaling laws governing Regime IV and V, remain elusive and warrant further investigation. Combining experimental, numerical, and theoretical efforts focused on pore-scale dynamics, boundary-layer behavior, and geometric and thermal parameter effects will be crucial to addressing these open questions.


\setcounter{table}{0}   
\setcounter{figure}{0}
\setcounter{section}{0}
\setcounter{equation}{0}
\renewcommand{\thetable}{A.\arabic{table}}
\renewcommand{\thefigure}{A.\arabic{figure}}
\renewcommand{\thesection}{A\arabic{section}}
\renewcommand{\theequation}{A\arabic{equation}}

\begin{appendix}

\section{The permeability $K$ and medium thermal conductivity $k_m$ of the porous media}\label{appA}
\begin{figure}[htbp]
    \centering
    \includegraphics[width=0.8\linewidth]{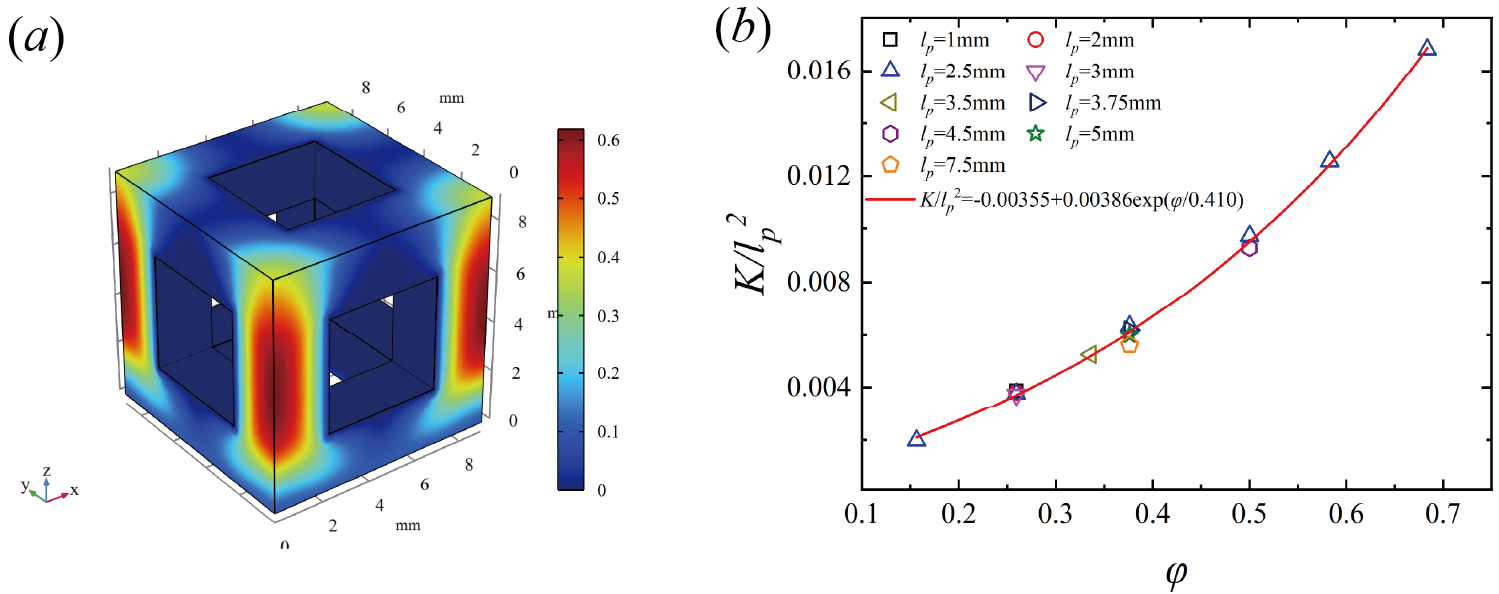}
    \caption{({\it a}) Velocity contour obtained from the simulation for $l=l_p=4.5 \rm mm$. ({\it b}) $K/l_p^2$ as a function of porosity $\varphi$.}
    \label{figK}
\end{figure}

\textcolor{black}{In the present study, the permeability $K$ and effective thermal conductivity $k_m$ of the porous medium were obtained using high-resolution 3D numerical simulations rather than direct experimental measurements. Accurate experimental determination of $K$ requires precise measurements of both pressure drop and flow rate, whereas quantifying $k_m$ requires high-precision heat flux measurements under pure conduction conditions. Such experimental requirements are challenging to fulfill with the current experimental setup. Considering the well-defined geometry of the 3D-printed porous structure employed in the present study, we can reliably determine $K$ and $k_m$ in a controlled and efficient manner by numerically solving the Stokes equation and the conduction equation, respectively.}

The simulations were performed using COMSOL Multiphysics 6.0. The computational domain was a cube with a side length of $L_{unit}=l+l_p$, which corresponds to one elementary unit of the porous structure (see figure~\ref{fig1} ({\it b})). The domain was discretised using a tetrahedral mesh, with local refinements applied near the fluid-solid interfaces to resolve the velocity and temperature gradients. 

To determine the permeability $K$, we solved the pressure-driven incompressible Stokes flow within the fluid phase. A constant pressure difference $\Delta p$ was imposed between the inlet and outlet along the flow direction ($z$-direction), and periodic boundary conditions were applied on the remaining four faces. The permeability $K$ was calculated as:
\begin{equation}
    K=u_{out}\mu_f\frac{L_{unit}}{\Delta p},
\end{equation}
where $u_{out}$ is the outlet velocity and $\mu_f$ is the dynamic viscosity of the fluid. Plots of $K/l_p^2$ versus porosity $\varphi$ for different pore dimensions are shown in figure~\ref{figK}, and these data points can be well fitted by an exponential function: 
\begin{equation}
    \frac{K}{l_p^2} = -0.00355+0.00386\rm exp^{\varphi/0.410}.
\end{equation}

The medium thermal conductivity, $k_m$, was determined by solving the steady-state heat conduction equation, with advection neglected. The computational domain was initialised with a constant temperature, and a constant heat flux was applied to the inlet. The other four faces were set to be adiabatic. The $k_m$ is given by:
\begin{equation}
    k_m = -\frac{Q}{\Delta T_s/L_{unit}},
\end{equation}
where $Q$ is the inlet heat flux and $\Delta T_s$ is the steady-state temperature difference across the domain. \textcolor{black}{The thermal conductivities of the solid phase and liquid phase used in the simulation are $k_s = 0.2~\mathrm{W/(m\cdot K)}$ and $k_f = 0.595~\mathrm{W/(m\cdot K)}$, which are consistent with the material properties stated in \S \ref{sec:Set-up}}. For the present porous medium, the simulation yields $k_m=0.344~\mathrm{W/(m\cdot K)}$, corresponding to a thermal conductivity ratio of $\lambda=1.75$.

\color{black}
\section{Determination of the critical Rayleigh--Darcy numbers}\label{appB}
\begin{figure}
    \centering
    \includegraphics[width=0.5\linewidth]{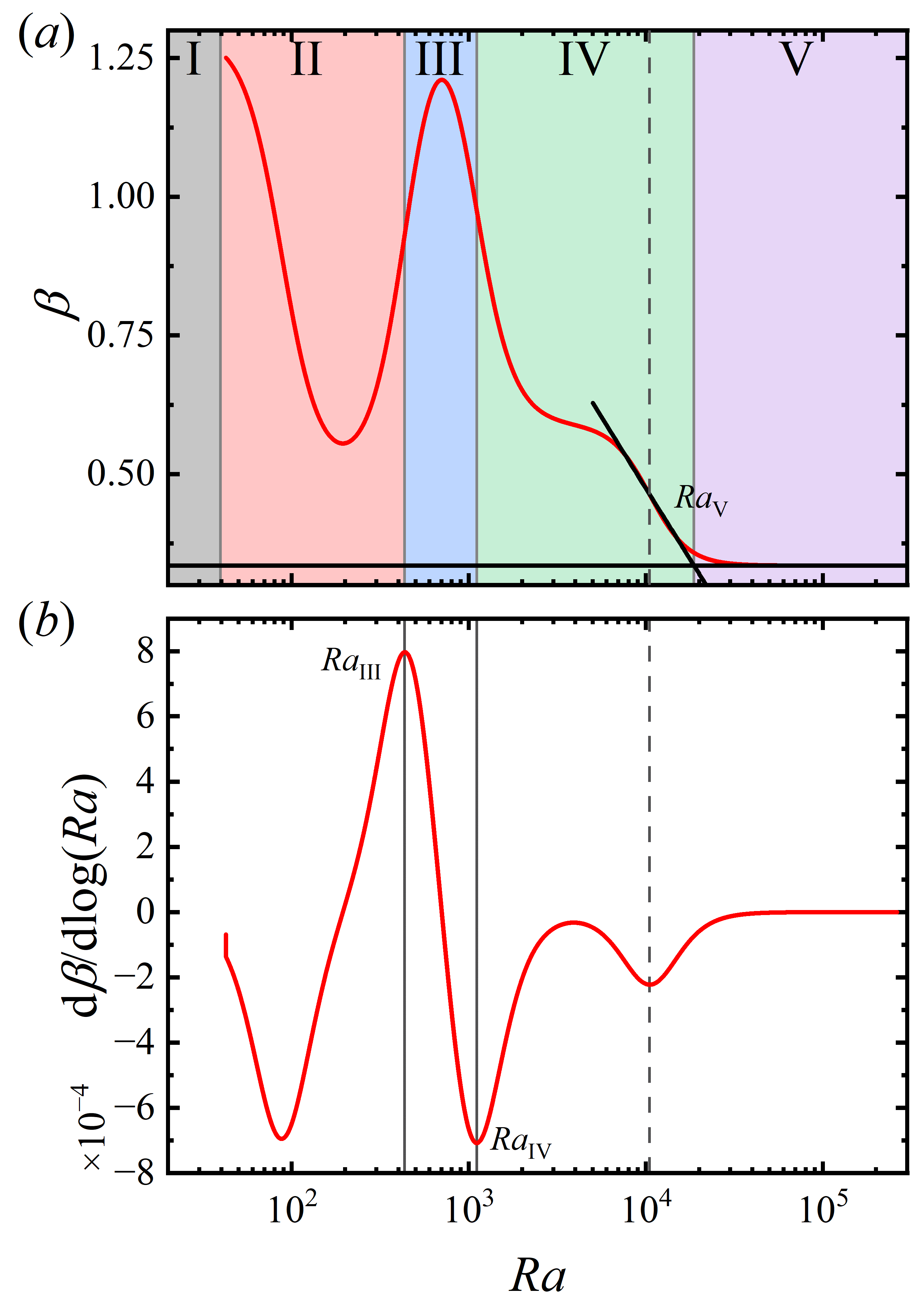}
    \caption{(a) Local scaling exponent $\beta$ of equation \ref{trans} for the dataset with $Da=1.21\times10^{-5}$, and (b) the corresponding gradient of $\beta$ as a function of the Rayleigh--Darcy number $Ra$.}
    \label{fig:Ra_cr}
\end{figure}

For the fitted function obtained from equation \ref{trans}, we first computed the local scaling exponent $\beta(Ra) = \rm{d} \log Nu_m / \rm{d} \log Ra$ (figure \ref{fig:Ra_cr}(a)). Its derivative $\rm{d}\beta / \rm{d}\log Ra$ was then evaluated, and local extrema were identified (see figure \ref{fig:Ra_cr}(b)). The $Ra$ values at the first local maximum and the second local minimum were taken as the critical Rayleigh--Darcy numbers $Ra_{\rm III}$ and $Ra_{\rm IV}$, respectively. Finally, the critical Rayleigh--Darcy number marking the onset of Regime V, $Ra_{\rm V}$, was determined as the intersection between the tangent to $\beta(Ra)$ at its last minimum with the horizontal line corresponding to the asymptotic scaling exponent $\beta_5$ (see figure \ref{fig:Ra_cr}(a)).
\color{black}

\section{Full experimental dataset}\label{appC}
The full experimental dataset of the present study is listed in table~\ref{Full_data}.

\newcommand{\casespacer}{\rule[-0.6ex]{0pt}{5ex}}
\begin{landscape}
\small
\setlength\LTleft{0pt}
\setlength\LTright{0pt}
\renewcommand{\arraystretch}{1.15}

\begin{longtable}{c|c|*{6}{>{\centering\arraybackslash}c >{\centering\arraybackslash}c|}>{\centering\arraybackslash}c >{\centering\arraybackslash}c}
\caption{Full experimental data.}\label{Full_data}\\
\hline
Case & $Da$ & $Ra$ & $Nu_m$  & $Ra$ & $Nu_m$  & $Ra$ & $Nu_m$  & $Ra$ & $Nu_m$  & $Ra$ & $Nu_m$  & $Ra$ & $Nu_m$  & $Ra$ & $Nu_m$ \\
\hline
\endfirsthead

\hline
Case & $Da$ & $Ra$ & $Nu_m$  & $Ra$ & $Nu_m$  & $Ra$ & $Nu_m$  & $Ra$ & $Nu_m$  & $Ra$ & $Nu_m$  & $Ra$ & $Nu_m$  & $Ra$ & $Nu_m$ \\
\hline
\endhead

\hline
\multicolumn{16}{r}{Continued on next page}\\
\endfoot

\hline
\endlastfoot

\multirow{3}{*}{PMC1} & \multirow{3}{*}{$1.21\times10^{-5}$} & \casespacer $1.07\times10^{3}$ & 19.0 & $8.47\times10^{2}$ & 15.2 & $6.72\times10^{2}$ & 11.8 & $5.34\times10^{2}$ & 9.45 & $4.25\times10^{2}$ & 7.71 & $3.36\times10^{2}$ & 6.74 & $2.68\times10^{2}$ & 5.97\\
& & $2.13\times10^{2}$ & 5.35 & $1.70\times10^{2}$ & 4.86 & $1.34\times10^{2}$ & 4.42 & $1.07\times10^{2}$ & 4.04 & $8.46\times10^{1}$ & 3.65 & $6.72\times10^{1}$ & 3.17 & $5.34\times10^{1}$ & 2.75\\
& & $4.26\times10^{1}$ & 2.33 & $3.36\times10^{1}$ & 2.24 & $2.68\times10^{1}$ & 2.16 & $1.33\times10^{3}$ & 22.9 & $1.29\times10^{3}$ & 22.1 &  &  &  & \\

\multirow{3}{*}{PMC2} & \multirow{3}{*}{$1.21\times10^{-5}$} & \casespacer $2.15\times10^{4}$ & 101.5 & $1.04\times10^{4}$ & 74.2 & $1.72\times10^{4}$ & 92.8 & $1.33\times10^{4}$ & 83.2 & $2.71\times10^{4}$ & 110.0 & $6.80\times10^{3}$ & 62.4 & $4.18\times10^{3}$ & 48.0\\
& & $5.44\times10^{3}$ & 55.9 & $2.70\times10^{3}$ & 36.7 & $3.42\times10^{3}$ & 42.8 & $8.60\times10^{3}$ & 69.7 & $3.43\times10^{4}$ & 115.7 & $1.06\times10^{3}$ & 18.7 & $1.37\times10^{3}$ & 23.9\\
& & $2.17\times10^{3}$ & 32.8 & $1.34\times10^{3}$ & 23.1 & $1.06\times10^{3}$ & 18.9 & $1.73\times10^{3}$ & 29.2 & $8.05\times10^{2}$ & 13.9 & $4.40\times10^{4}$ & 127.6 & $3.34\times10^{4}$ & 116.3 \\

\multirow{3}{*}{PMC3} & \multirow{3}{*}{$1.21\times10^{-5}$} & \casespacer $1.71\times10^{5}$ & 204.6 & $2.14\times10^{5}$ & 218.6 & $1.35\times10^{5}$ & 187.7 & $1.04\times10^{5}$ & 173.4 & $6.76\times10^{4}$ & 151.9 & $8.47\times10^{4}$ & 163.9 & $5.31\times10^{4}$ & 140.5\\
& & $4.27\times10^{4}$ & 130.1 & $2.13\times10^{4}$ & 102.8 & $2.69\times10^{4}$ & 113.5 & $1.70\times10^{4}$ & 94.8 & $1.05\times10^{4}$ & 76.6 & $1.38\times10^{4}$ & 87.9 & $3.37\times10^{4}$ & 122.8\\
& & $2.63\times10^{5}$ & 232.0 &  &  &  &  &  &  &  &  &  &  &  &  \\

\multirow{3}{*}{PMC4} & \multirow{3}{*}{$5.39\times10^{-6}$} & \casespacer $1.89\times10^{3}$ & 33.5 & $1.49\times10^{3}$ & 25.8 & $1.19\times10^{3}$ & 19.9 & $9.45\times10^{2}$ & 15.4 & $7.54\times10^{2}$ & 12.4 & $5.97\times10^{2}$ & 10.6 & $4.74\times10^{2}$ & 8.88\\
& & $3.76\times10^{2}$ & 8.14 & $2.99\times10^{2}$ & 7.49 & $2.38\times10^{2}$ & 6.61 & $1.89\times10^{2}$ & 5.85 & $1.50\times10^{2}$ & 5.14 & $1.19\times10^{2}$ & 4.56 & $9.45\times10^{1}$ & 4.10\\
& & $7.55\times10^{1}$ & 3.66 & $5.97\times10^{2}$ & 10.4 & $4.74\times10^{2}$ & 9.24 &  &  &  &  &  &  &  & \\

\multirow{4}{*}{PMC5} & \multirow{4}{*}{$5.39\times10^{-6}$} & \casespacer $2.70\times10^{3}$ & 39.1 & $1.72\times10^{3}$ & 25.9 & $2.13\times10^{3}$ & 32.3 & $1.04\times10^{3}$ & 14.7 & $1.32\times10^{3}$ & 18.5 & $5.97\times10^{3}$ & 67.6 & $7.55\times10^{3}$ & 76.3\\
& & $9.46\times10^{3}$ & 86.2 & $1.19\times10^{4}$ & 96.6 & $1.47\times10^{4}$ & 107.0 & $4.74\times10^{3}$ & 58.7 & $3.74\times10^{3}$ & 49.8 & $2.99\times10^{3}$ & 41.8 & $2.37\times10^{3}$ & 36.6\\
& & $1.90\times10^{3}$ & 29.5 & $1.49\times10^{3}$ & 22.6 & $1.19\times10^{3}$ & 16.8 & $9.57\times10^{2}$ & 13.9 & $7.52\times10^{2}$ & 12.0 & $5.80\times10^{2}$ & 10.8 & $1.20\times10^{3}$ & 16.8\\
& & $2.38\times10^{3}$ & 35.7 & $1.89\times10^{3}$ & 28.9 & $1.51\times10^{3}$ & 21.8 &  &  &  &  &  &  &  & \\

\multirow{3}{*}{PMC6} & \multirow{3}{*}{$5.39\times10^{-6}$} & \casespacer $4.73\times10^{3}$ & 53.2 & $7.54\times10^{3}$ & 73.3 & $1.19\times10^{4}$ & 92.4 & $1.50\times10^{4}$ & 104.3 & $1.89\times10^{4}$ & 111.0 & $2.37\times10^{4}$ & 128.8 & $2.99\times10^{4}$ & 138.7\\
& & $3.76\times10^{4}$ & 155.5 & $4.74\times10^{4}$ & 170.4 & $5.97\times10^{4}$ & 184.1 & $7.54\times10^{4}$ & 201.7 & $9.45\times10^{4}$ & 216.8 & $1.19\times10^{5}$ & 232.1 & $1.49\times10^{5}$ & 250.8\\
& & $4.71\times10^{3}$ & 50.4 & $1.91\times10^{4}$ & 110.1 &  &  &  &  &  &  &  &  &  & \\

\multirow{3}{*}{PMC7} & \multirow{3}{*}{$2.48\times10^{-6}$} & \casespacer $4.34\times10^{3}$ & 59.3 & $3.46\times10^{3}$ & 49.6 & $2.74\times10^{3}$ & 39.3 & $2.17\times10^{3}$ & 31.2 & $1.73\times10^{3}$ & 23.0 & $1.37\times10^{3}$ & 18.2 & $1.09\times10^{3}$ & 15.7\\
& & $8.67\times10^{2}$ & 14.1 & $6.86\times10^{2}$ & 12.5 & $5.47\times10^{2}$ & 10.9 & $4.34\times10^{2}$ & 10.1 & $5.47\times10^{2}$ & 65.3 & $6.86\times10^{2}$ & 81.1 & $5.47\times10^{2}$ & 68.3\\
& & $4.37\times10^{3}$ & 56.9 &  &  &  &  &  &  &  &  &  &  &  & \\

\multirow{2}{*}{PMC8} & \multirow{2}{*}{$1.35\times10^{-6}$} & \casespacer $3.73\times10^{3}$ & 55.8 & $2.98\times10^{3}$ & 44.2 & $2.36\times10^{3}$ & 33.8 & $1.89\times10^{3}$ & 26.1 & $1.49\times10^{3}$ & 20.7 & $1.18\times10^{3}$ & 18.1 & $9.41\times10^{2}$ & 16.9\\
& & $7.47\times10^{2}$ & 16.0 & $5.93\times10^{2}$ & 14.3 & $4.73\times10^{2}$ & 13.9 &  &  &  &  &  &  &  & \\

\multirow{3}{*}{PMC9} & \multirow{3}{*}{$1.35\times10^{-6}$} & \casespacer $1.89\times10^{3}$ & 25.2 & $2.98\times10^{3}$ & 40.9 & $4.73\times10^{3}$ & 65.7 & $5.93\times10^{3}$ & 77.4 & $3.74\times10^{3}$ & 51.8 & $2.36\times10^{3}$ & 31.8 & $1.49\times10^{3}$ & 21.5\\
& & $1.16\times10^{3}$ & 19.2 & $1.47\times10^{3}$ & 21.8 & $1.88\times10^{3}$ & 25.7 & $2.39\times10^{3}$ & 32.2 & $8.79\times10^{2}$ & 17.8 & $7.47\times10^{3}$ & 91.4 & $1.18\times10^{4}$ & 124.4\\
& & $1.49\times10^{4}$ & 147.7 & $1.85\times10^{4}$ & 164.1 & $2.36\times10^{4}$ & 183.2 & $2.98\times10^{4}$ & 202.3 & $9.61\times10^{3}$ & 109.1 &  &  &  & \\

\multirow{3}{*}{PMC10} & \multirow{3}{*}{$6.18\times10^{-7}$} & \casespacer $1.03\times10^{3}$ & 18.0 & $1.37\times10^{3}$ & 21.6 & $1.72\times10^{3}$ & 25.1 & $2.17\times10^{3}$ & 28.6 & $2.72\times10^{3}$ & 32.5 & $3.41\times10^{3}$ & 38.9 & $4.32\times10^{3}$ & 50.6\\
& & $5.43\times10^{3}$ & 65.1 & $1.35\times10^{4}$ & 137.2 & $1.09\times10^{4}$ & 115.2 & $8.64\times10^{3}$ & 97.0 & $7.54\times10^{3}$ & 92.0 & $6.12\times10^{3}$ & 75.3 & $6.84\times10^{3}$ & 84.0\\
& & $8.72\times10^{3}$ & 99.2 & $1.71\times10^{4}$ & 160.3 &  &  &  &  &  &  &  &  &  & \\

\multirow{3}{*}{RBC1} & \multirow{3}{*}{-} & \casespacer $6.31 \times 10^7$ & 27.8 & $3.98 \times 10^7$ & 24.4 & $2.00 \times 10^7$ & 20.1 & $1.00 \times 10^7$ & 16.6 & $6.33 \times 10^6$ & 14.6 & $3.98 \times 10^6$ & 12.8 & $2.54 \times 10^6$ & 11.4\\
& & $1.58 \times 10^6$ & 10.1 & $1.26 \times 10^6$ & 9.46 & $2.00 \times 10^6$ & 10.7 & $5.02 \times 10^6$ & 13.7 & $3.16 \times 10^6$ & 12.1 & $7.98 \times 10^6$ & 15.6 & $1.26 \times 10^7$ & 17.7\\
& & $1.58 \times 10^7$ & 18.9 & $2.51 \times 10^7$ & 21.5 & $3.16 \times 10^7$ & 22.9 & $5.01 \times 10^7$ & 26.1 &  &  &  &  &  & \\

\multirow{2}{*}{RBC2} & \multirow{2}{*}{-} & \casespacer $1.25 \times 10^8$ & 33.9 & $2.00 \times 10^8$ & 38.8 & $8.05 \times 10^7$ & 30.2 & $6.29 \times 10^7$ & 28.1 & $4.99 \times 10^7$ & 26.3 & $3.13 \times 10^7$ & 23.1 & $1.98 \times 10^7$ & 20.2\\
& & $9.83 \times 10^6$ & 16.7 & $1.27 \times 10^7$ & 17.8 & $1.57 \times 10^7$ & 18.9 & $3.99 \times 10^7$ & 24.6 & $2.49 \times 10^7$ & 21.5 & $1.00 \times 10^8$ & 31.9 & $1.58 \times 10^8$ & 36.3 \\

\multirow{3}{*}{RBC3} & \multirow{3}{*}{-} & \casespacer $1.02 \times 10^9$ & 65.6 & $1.26 \times 10^9$ & 69.9 & $1.58 \times 10^9$ & 75.0 & $7.73 \times 10^8$ & 59.9 & $6.26 \times 10^8$ & 56.2 & $4.96 \times 10^8$ & 52.3 & $3.94 \times 10^8$ & 49.0\\
& & $3.15 \times 10^8$ & 45.8 & $2.42 \times 10^8$ & 42.2 & $1.42 \times 10^8$ & 36.0 & $1.27 \times 10^8$ & 35.3 & $9.46 \times 10^7$ & 32.2 & $2.10 \times 10^8$ & 41.3 & $2.00 \times 10^9$ & 82.3\\
& & $9.71 \times 10^8$ & 64.8 &  &  &  &  &  &  &  &  &  &  &  & \\

\multirow{2}{*}{RBC4} & \multirow{2}{*}{-} & \casespacer $5.11 \times 10^9$ & 110.0 & $6.35 \times 10^9$ & 116.9 & $1.02 \times 10^{10}$ & 135.2 & $7.98 \times 10^9$ & 125.3 & $3.07 \times 10^9$ & 92.8 & $4.01 \times 10^9$ & 101.0 & $1.95 \times 10^9$ & 82.5\\
& & $2.54 \times 10^9$ & 89.3 & $9.49 \times 10^8$ & 67.6 & $1.62 \times 10^9$ & 79.1 & $1.29 \times 10^{10}$ & 145.6 & $1.60 \times 10^{10}$ & 155.3 &  &  &  & \\
\end{longtable}
\end{landscape}

\end{appendix}
\clearpage

\begin{bmhead}[Acknowledgements]
We are grateful to Dr. De Paoli, Mr. V. Giurgiu, and Dr. K. Xu for fruitful discussions.
\end{bmhead}

\begin{bmhead}[Funding]
This work is supported by the National Natural Science Foundation of China (Grants No. 12202173, No. 12232010, No. 12595300 and No. 12595302).
\end{bmhead}

\begin{bmhead}[Declaration of interests]
The authors report no conflict of interest. 
\end{bmhead}

\begin{bmhead}[Data availability statement]
The data of this study are available from the corresponding author upon request.
\end{bmhead}

\begin{bmhead}[Author ORCIDs]
 J. Dong, https://orcid.org/0000-0002-7943-9862; L. Zhang, https://orcid.org/0000-0003-4009-2969; K.-Q. Xia, https://orcid.org/0000-0001-5093-9014.
\end{bmhead}

\begin{bmhead}[Author contributions]
J. Dong and L. Zhang contributed equally to this work. J. Dong conducted the experiments, performed data analysis, and drafted the manuscript; L. Zhang conceived the project, performed data analysis, and revised the manuscript; K.-Q. Xia performed data analysis, revised the manuscript and supervised the project. 
\end{bmhead}

\bibliographystyle{jfm}
\bibliography{jfm}

\end{document}